\documentstyle[12pt,epsfig,amssymb]{article}
\begin{document}
\begin{flushright}
CERN-TH/97-289 \\
\end{flushright}

\begin{center}
{\LARGE\bf NLO Analysis of Semi-inclusive DIS}
\vspace{0.6cm}

\renewcommand{\thefootnote}{\fnsymbol{footnote}}
\setcounter{footnote}{0}
{\Large Daniel de Florian\footnote{Invited talk presented at the workshop 
`Deep Inelastic Scattering off Polarized Targets:  Theory Meets Experiment', 
Zeuthen, Germany, September 1-5, 1997}}
\renewcommand{\thefootnote}{\fnsymbol{roman}}
\setcounter{footnote}{0}

\vspace*{0.4cm}
{\it Theory Division, CERN, CH-1211 Geneva, 
Switzerland}\\

\vspace*{0.3cm}

\end{center}

\begin{abstract}
 We present a combined NLO QCD analysis to data on both inclusive and semi-inclusive polarized asymmetries. We also present the NLO corrections to the direct part of the polarized photoproduction of charged hadrons.
 
\end{abstract}
\vspace*{-0.4cm}

\section{Introduction}
In recent years, considerable attention has been paid to polarized deep 
inelastic scattering experiments, to the interpretation of the corresponding
data in the framework of perturbative QCD, and to the phenomenological
extraction of non-perturbative spin-dependent parton distributions.
 
The intense activity around these issues have come not only from the
interesting developments and discussions that have arisen in
each of them, but also from the fact that, combined, they are the most
appropriate tools to unveil the
spin structure of nucleons, a subject that is still being debated.

In fact, an increasing amount of high-precission totally inclusive
data, collected by different collaborations over the last few years
combined with the recent computation of the complete
perturbative QCD corrections up to next-to-leading order of
the inclusive cross sections , have lead to
several QCD analyses and also extractions of polarized parton
distributions \cite{fits}.  They demostrate that these data are not sufficient to accurately extract the spin-dependent quark and gluon densities of the nucleon.
In order to improve our knowledge on polarized parton distributions new 
less-inclusive processes than DIS have to be considered.

One of the sources foreseen for additional data that can be included in
those analyses is the so-called semi-inclusive spin-dependent
asymmetries. These asymmetries are particularly sensitive to specific
combinations of partons of different flavours and have been
proposed and used to study the valence-quark distributions in the
proton \cite{jorg}.  
More recently, a large amount of more accurate semi-inclusive data have
been produced, and also the appropriate perturbative tools for their
analysis have been developed \cite{npb1}.
In the first part of this work, we  show the results of a NLO extraction of polarized parton distributions from both inclusive and semi-inclusive data and analyze the impact of the semi-inclusive one in the global fit, including the obtention of   {\it valence} distributions from the semi-inclusive data alone \cite{fitnos}.

On the other hand, in order to obtain information on the polarized gluon distribution it is necessary to pick up an observable which has a gluonic contribution already at lowest order. One of them is the production of a (charged) hadron with large transverse momentum $p_T$, specially in the case of photoproduction where a larger number of hadrons is obtained. Compared to the case of jet-production, the production of charged hadrons allows to go to lower values of $p_T$. It has been shown in \cite{werner} that a polarized version of HERA collider would give a very promising and useful facility to study photoproduction reactions.
In order to make reliable quantitative predictions for such a high-energy proces, it is crucial to extend LO studies to NLO. The key issues here are to check the perturbative stability  and to partially cancell the scale dependence of the observable. As the second part of this contribution we show the results obtained from a NLO computation of the {\it direct} part of the (semi-inclusive) photoproduction of hadrons \cite{danielwerner}.

\vspace{1mm}
\noindent

\section{Semi-Inclusive Asymmetries}

The semi-inclusive asymmetry  can be written as: 
 \begin{equation} \left.
A_{1}^{N\,h}(x,Q^2) \right|_Z \simeq \frac{\int_{Z} dz\,
g_{1}^{N\,h}(x,z,Q^2)}{\int_{Z} dz\,F_{1}^{N\,h}(x,z,Q^2)} ,
\end{equation} 
where the superscript $h$ denotes the hadron detected in
the final state, and  the variable $z$ is given by the ratio between the hadron
energy and that of the spectators in the target.

The semi-inclusive spin-dependent structure function
$g_{1}^{N\,h}(x,z,Q^2)$ can  be decomposed into convolutions
between parton densities $\Delta q_i(x,Q^2)$, $\Delta g(x,Q^2)$, unpolarized fragmentation
functions $D_{h/j}(z,Q^2)$,  coefficient functions $\Delta C_{ij}$, and
polarized fracture functions $\Delta M^h_{i}(x,z,Q^2)$, the latter being given by the 
contribution to the target fragmentation region as
\begin{eqnarray} 
g_{1}^{N\,h}(x,z,Q^2) &=& \sum_{q,\overline{q}} c_i \left
\{ \Delta q_i (x,Q^2)  D_{h/i}(z,Q^2) +\frac{\alpha_s(Q^2)}{2\pi}[ \Delta q_i
 \otimes \Delta C_{ij} \otimes D_{h/j} \right.  \nonumber \\
&+& \left. \Delta
q_i  \otimes \Delta C_{ig} \otimes D_{h/g}+ \Delta g  \otimes
\Delta C_{gj} \otimes D_{h/j} ] \right.  \\ 
&+& \left. \Delta M_{q_i}^h (x,z,Q^2)
+\frac{\alpha_s(Q^2)}{2\pi}[\Delta M_{q_i}^h \otimes \Delta C_{i}+\Delta
M_{g}^h \otimes \Delta C_{g}] \right \} \nonumber .
\end{eqnarray} 
A complete computation
of this kind of observable and the full expressions for the
corresponding coefficient functions in different factorization schemes
can be found in ref. \cite{npb1}.  An analogous expression can be
written for the unpolarized semi-inclusive structure function
\cite{graudenz}.

  In our computations we use the
charged pion and kaon fragmentation functions of ref. \cite{fragmen}  combined with a parametrization of semi-inclusive EMC data
\cite{emc}.
  The unpolarized observables are constructed  using the parton distributions of ref. \cite{grv} in their LO and NLO
($\overline{MS}$) versions, according to the order of the fit, and with the 
appropiate QCD coefficients.
Polarized and unpolarized fracture functions \cite{veneziano,graudenz,npb1,shore}
describe the details of hadro\-ni\-za\-tion processes coming mainly
from target fragmentation region. Although their inclusion is
crucial in order to consistently factorize collinear divergences, once
this process is through, their actual contribution to the cross sections
can be be suppressed by imposing the appropriate kinematical cuts.
  Consequently, we
restrict our analysis to single asymmetries for $z_h>0.2$,
leaving for the moment the discussion of difference asymmetries, and
neglecting fracture function contributions.

In the present analysis, rather
than adopting some or other stringent constraint on the normalization
of the valence, sea quarks, or gluon densities,  then singling out
the set that presents the lowest $\chi^2$ (given those and other less
apparent assumptions), we adopt a more flexible scheme for the valence
and sea sectors, we put greater emphasis on the measured region, and we
explore different gluon possibilities.  

As we are primarily interested in the measured region, we adopt a rather
simple parametric form for the input spin-dependent valence quark
densities: 
 \begin{equation}
 x \Delta q_V (x,Q_0^2)=N_{q_V}
\frac{x^{\alpha_q}(1-x)^{\beta_q}(1+\gamma_q\,
x)}{B(\alpha_q+1,\beta_q+1)+\gamma_q \, B(\alpha_q+2,\beta_q+1) },
\end{equation}
 where the parameters $\alpha_q$ and $\gamma_q$ are
obtained from the fitting procedure, and $\beta_q$ is externally fixed
by the positivity constraint with respect to GRV unpolarized parton
distributions at large $x$. 
The initial scale $Q^2_0$ is chosen to be
$0.5\, $GeV$^2$. In order to trace and parametrize the departure from the SU(2)
and SU(3) flavour symmetries, we define the normalization
coefficients $N_{q_V}$ in terms of the $F$ and $D$ constants and two
additional  parameters.
\begin{equation} 
\delta u_V - \delta
d_V = (F+D)(1+\epsilon_{Bj}) 
\end{equation} 
and 
\begin{equation}
 \delta u_V + \delta d_V  + 4 (\delta \overline{u}-\delta s ) =
(3F-D)(1+\epsilon_{SU(3)}).
 \end{equation}
 The parameters
$\epsilon_{Bj}$ and $\epsilon_{SU(3)}$ account quantitatively for
eventual departures from  flavour symmetry considerations (including
also some uncertainties on the low-$x$ behaviour). They also measure the
degree of fulfilment of the Bjorken  and Ellis-Jaffe sum rules.
For the light quarks the proposed input density is given by:
\begin{equation}
 x\Delta \overline{q} (x,Q_0^2)=N_{\overline{q}}
\frac{x^{\alpha_{\overline{q}}}(1-x)^{\beta_{\overline{q}}}}
{B(\alpha_{\overline{q}}+1,\beta_{\overline{q}}+1)},
\end{equation}
 where  $\alpha_{\overline{q}}$, $\beta_{\overline{q}}$,
and $N_{\overline{q}}$ are  only constrained by
positivity. The same functional dependence and considerations are used
for gluons, since using more pa\-ra\-me\-ters seems to be useless, taking into 
account the uncertainties on them. For strange quarks we adopt: 
 \begin{equation}
 \Delta \overline{s} (x,Q_0^2)=N_{\overline{s}}\, \Delta \overline{q} (x,Q_0^2),
\end{equation} 
finding pointless the addition of more parameters.
  In order to avoid possible higher-twist contributions, we have
taken into account only measurements with $Q^2>1\,$GeV$^2$ given a total of 133 inclusive data points. As
semi-inclusive data we take those recently presented by SMC \cite{jorg}, 48
data points, which then lead to combined global fits with 181 data points.
Correlations between totally-inclusive and semi-inclusive SMC data sets
have been taken into account, and increase the total $\chi^2$.
\begin{center}
\begin{tabular}{|c|c|c|c|c|c|c|} \hline \hline
      &  \multicolumn{3}{c|}{NLO {\footnotesize ($\overline{MS}$)} }&
   \multicolumn{3}{c|}{LO} \\ \cline{2-7}
    & {\footnotesize Set 1}  & {\footnotesize Set 2} &{\footnotesize
    Set 3}  & {\footnotesize  Set 1 } & {\footnotesize  Set
2} & {\footnotesize Set 3 }\\ \hline\hline $
 \chi^2_{T}$&151.10 & 149.89  & 150.10 & 155.86 & 154.80 & 156.99\\ \hline 
$\chi^2_{I}$&101.90 & 100.47        & 100.84 & 107.56 &  106.37      & 108.73 \\ \hline
$\chi^2_{SI}$&44.02 &  46.03     &  46.15 & 45.19  & 45.33    & 44.91  \\ \hline\hline
\end{tabular}
\end{center} 
\begin{center}
 {\bf Table 1: $\chi^2$ values for total ($T$), inclusive $(I)$ and semi-inclusive $(SI)$ data. }
\end{center}

In Table 1 we show the the best $\chi^2$ values obtained for three different NLO ($\overline{MS}$)
 and LO gobal fits for combined inclusive and
semi-inclusive data in which the gluon density first moments $N_g$ are
constrained  to three different regions:
\begin{eqnarray}
\rm{Set\, 1} &&\,\,\,\,\,\,\,\,\,\,\,\,\,\,\,\,\, \delta g > 0.8\, \nonumber \\
\rm{Set\, 2} &&\,\, 0.1 > \delta g  > 0.8 \nonumber \\
\rm{Set\, 3} &&\,\,\,\,\,\,\,\,\,\,\,\,\,\,\,\,\,\delta g < 0.1 , \nonumber 
\end{eqnarray}
 defined at the initial scale. The breaking parameter
 $\epsilon_{Bj}$  is left free whereas,   $\epsilon_{SU(3)}$  is constrained to allow only
 moderate violations of the polarized sum rules.

 Clearly, the semi-inclusive data set is in  very good agreement  with the inclusive one, and allows fits of remarkable quality in the three gluon regions.
In the combined fits there is a   preference for sets with a   moderate gluon polarization (set 2), as found in other analysis.
However, the differences in $\chi^2$ values obtained in each of the regions are so subtle that the uncertainty in the value for the first moment of the polarized gluon density is significantly large, and even a slightly negatively polarized distribution for gluons can not be ruled out yet.

In fig. 1 we compare the inclusive asymmetries coming from our best Set 2 (NLO and LO, respectively) with the data, other sets give very similar results. 
\vspace*{-1cm}
\begin{figure}[htb]
\begin{center}
\mbox{\kern-1cm
\epsfig{file=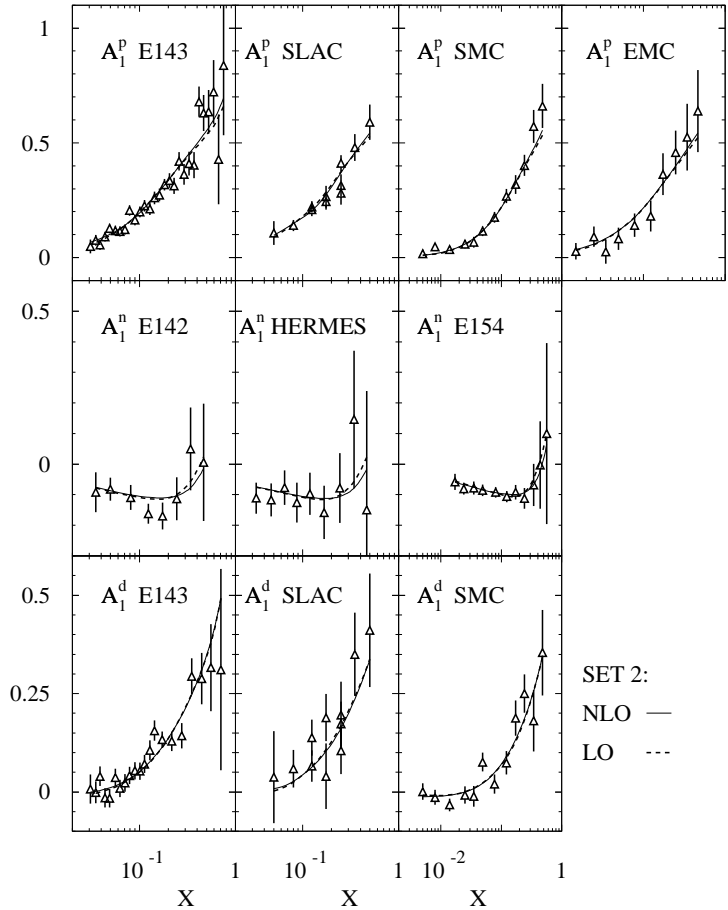,width=11.0truecm,angle=0}}
\caption{ Inclusive asymmetry data against the expectations coming from Set 2.}
\end{center}
\end{figure}

In fig. 2 we show the same but for the semi-inclusive data. Notice that the large error bars of these data reduce its weight in the global fit and that the main difference in the $\chi^2$ between LO and NLO fits comes from the totally inclusive data. Also in fig. 2 we show the result of a fit using only the semi-inclusive data as described below. 
\begin{figure}[htb]
\begin{center}
\mbox{\kern-1cm
\epsfig{file=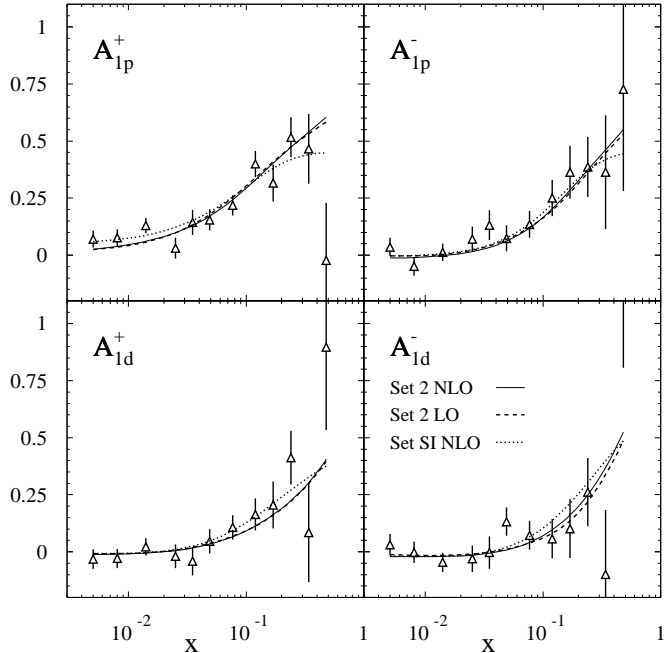,width=10.0truecm,angle=0}}
\caption{
The same as Fig. 1, but for semi-inclusive asymmetries,
and the expectation coming from the semi-inclusive set (dots).}
\end{center}
\end{figure}

In Table 2   we show sum rules and first moments estimates  for Set 2 at different scales. For the Bjorken sum rule $\Gamma^{Bj}$, the departure from the theroretical expectation is significantly small, as given by the small values found for the parameter $\epsilon_{Bj}$. 
 
As usual in the $\overline{MS}$ scheme,
the first moment of the singlet distribution, $\delta \Sigma$, is found to be considerably smaller than the naive prediction, and is correlated to the gluon polarization. Notice that the valence-quark normalizations are quite stable and give the same result, independently of the singlet sector and that in the case of the polarized sea we show the first moment corresponding to $u$ and $d$ quarks, being negligible the differences with the one for $s$ quarks.
  
\begin{center}
\begin{tabular}{|c|c|c|c|c|c|c|c|c|c|} \hline \hline
{\footnotesize Fit} &  {\footnotesize $Q^2$}  & $\Gamma_1^p$ & $\Gamma_1^n$ & $\Gamma^{Bj}$ & $\delta\Sigma$ & $\delta g$ & $\delta u_V$ & $\delta d_V$ & $\delta \overline{q}$ \\ \hline \hline
 Set 2 & 1 & 0.124 & $-$0.057& 0.182 & 0.212 & 0.59& 0.875 & $-$0.354 & $-$0.051\\ \cline{2-10}
      & 4 & 0.129 & $-$0.060& 0.189 & 0.207 & 0.91& 0.873 & $-$0.354 & $-$0.052\\ \cline{2-10}
     & 10 & 0.130 & $-$0.061& 0.191 & 0.206 & 1.11& 0.873 & $-$0.354 & $-$0.052\\ \hline 
  \hline
\end{tabular} 

\vspace*{5mm} {\bf Table 2: Sum rules from a NLO combined fit}. \end{center}

The impact of the semi-inclusive data in the total fit has been estimated  performing also fits using only inclusive data. In these fits we have found that the quark parameters  change less than $2\%$, whereas the changes are a somewhat larger for the gluon distribution. However, the uncertainties already pointed
out about the gluon density dominate over any potential influence of the semi-inclusive data set. The reasons for this very small impact are, basically, the fact that semi-inclusive data has not reached yet the precision and statistical significance of the inclusive one, and also  that the data sets are not completely independent. This can be seen either in the correlations between inclusive and semi-inclusive asymmetries \cite{jorg}, and also in the fact that parametrizations obtained using only inclusive data give a very good description of the semi-inclusive asymmetries.    

Additionaly, it is possible to use the semi-inclusive data in QCD global fits but without employing the inclusive data sets directly, for the comparison of the corresponding results.  As in this case, not all the parameters can be unambiguously fixed by the semi-inclusive data alone,  we have fixed the ones corresponding to the gluon and sea densities to the values obtained in Set 2, and then adjusted only the valence-quark distributions. 

In these fits,
the $\chi^2$ values with respect to the semi-inclusive data, $\chi^2_{SI}$, are reduced in some units; however, the total $\chi^2$ increases dramatically to unacceptable values ($\chi^2_{T}>290$), the main difference being in the
$\Delta d_V$ distribution, as  can be seen in fig. 3, where the parton densities   are shown at the common value of  $Q^2=10$ GeV$^2$.  As can be observed the  $\Delta u_V$ distribution is in very good agreement with the one obtained in the global analysis.
\vspace*{-1cm}
 \begin{figure}[htb]
\begin{center}
\mbox{\kern-1cm
\epsfig{file=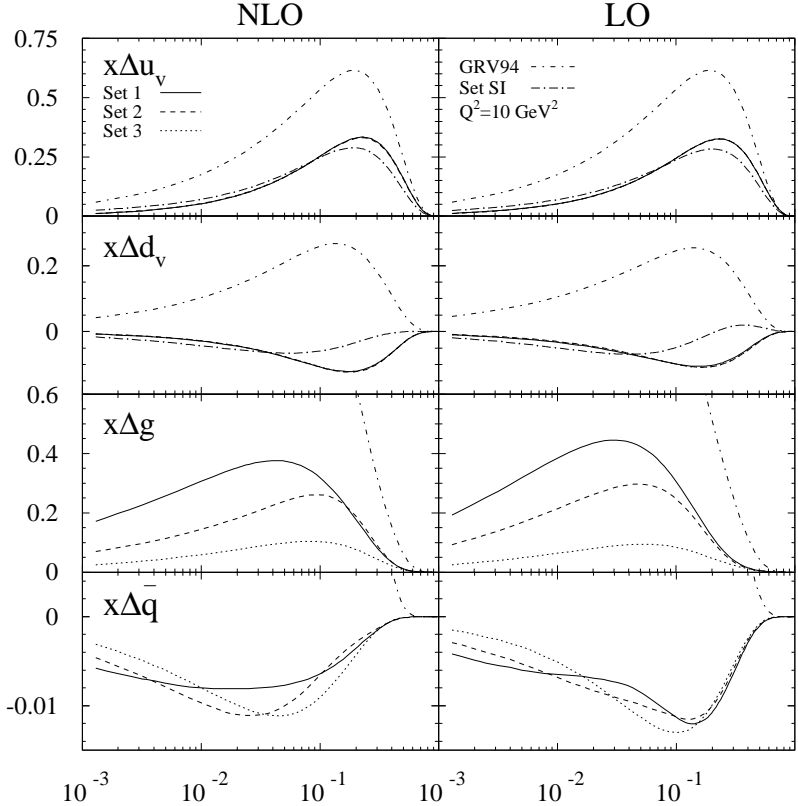,width=12.0truecm,angle=0}}
\caption{ 
Parton densities at 10 GeV$^2$.}
\end{center}
\end{figure}

In the semi-inclusive case, the $\Delta d_V$ distribution is mainly
   constrained by the  deuteron  asymmetry, different from the inclusive case, where  is determined by the more accurate E-154 neutron data. As  can be seen in fig. 2, the difference between  the result for the deuteron asymmetry coming either from the combined fit or  the semi-inclusive one is   apparent, 
even though the  $\Delta d_V$ distributions are quite different, showing the low sensitivity of deuteron observables to this density.   Of course all the distributions  are compatible if the errors not shown in the plot are taken into account. 
 
 Ongoing semi-inclusive measurements using $^3$He targets at HERMES
 can be quite useful
in the determination of valence-quark distributions from semi-inclusive data alone, and also as further constraints in global fits. These asymmetries are particularly sensitive to $\Delta d_V$, specially the one for the production of positively charged hadrons, as can be expected from very simple arguments based on the values of the corresponding fragmentation functions.   \\

\section{Hadron Photoproduction in NLO}

As it was already stated, the photoproduction of charged hadrons is a very well suited process to pin down  the polarized gluon distribution of the proton but also  the polarized parton distributions of the {\it photon} in the case of colliders.

The knowledge of the NLO corrections for this process allows to analyze the perturbative stability by means of the $K$-factor, which is essential in order to establish the kinematical range of applicability of the QCD method and to confirm the LO expectations \cite{werner}. It can tell us, for instance, which is the minimun value of $p_T$ that can be considered for this observable in order to increase the statistics but, at the same time, allowing  for a perturbative interpretation of the result. Moreover, at LO three different scales appear, $\mu$ as the argument of $\alpha_s$ and $M, M_F$ as the scales  where the parton distributions and the fragmentation functions are evaluated respectively. The LO treatment leads then to a strong dependence of the cross section  on the choice of these arbitrary scales, via either the value of the coupling constant or the Altarelli-Parisi evolution of the distributions. This dependence is partially cancelled by the addition of the NLO corrections. 
 
As it has been well established in the unpolarized case, the real photon will not only interact in a direct way, but can also be resolved into its hadronic structure. As was shown in \cite{werner}, the resolved component is 
subdominant with respect to the direct one in certain regions of 
rapidity and transverse momentum of the produced hadron or jet, thus 
maintaining the clear-cut sensitivity to $\Delta g$ resulting from the 
direct piece.  

 The first basic ingredient for the  extension
to NLO has been provided in the past two years by the fact mentioned in the last section that NLO
fits to polarized DIS data have been performed, yielding spin-dependent nucleon
parton distributions evolved to NLO accuracy. Focussing on the direct part 
of inclusive-hadron photoproduction, the calculation of the polarized
cross section to NLO is then completed by using also (unpolarized) NLO fragmentation 
functions for the produced hadron (as provided in \cite{fragmen}), and by including 
the ${\cal O} (\alpha_s)$ corrections to the spin-dependent direct subprocesses
for the inclusive production of a certain parton that fragments into
the hadron. The calculation of the latter is the purpose of this paper.
An immediate problem arises here as the direct part on its own is no longer 
a really well-defined quantity beyond the LO. 
This is due to the fact that beyond LO collinear singularities appear 
in the calculation of the subprocess cross sections for photon-parton scattering
which are to be attributed to a collinear spliting of the photon into a 
$q\bar{q}$ pair and need to be absorbed into the photon structure functions. 
As the latter only appear in the resolved part of the cross section, and since 
factorizing singularities is never a unique procedure, it follows that only the sum 
of the direct and the resolved pieces is independent of the factorization scheme 
chosen and thus physical. This has been known for a long time from the 
unpolarized case where the corrections to the direct \cite{lio} {\em and} to the 
resolved \cite{guil} contributions have all been calculated. Nevertheless, we will 
concentrate in this work only on the corrections to the direct part of the 
polarized cross section, mainly because this calculation -- albeit 
already being quite involved -- is much simpler than the one for the resolved piece. 
Our results will therefore only be the first step in a full calculation of NLO 
effects to polarized inclusive-hadron photoproduction. Despite the fact that they 
are not complete in the sense discussed above, we believe our results to be very 
important both phenomenologically and theoretically: 
the direct component dominates at fixed target energies and also still for the HERA 
collider situation in certain regions of phase space. This means that our 
NLO results should be rather close to the true NLO answer in these cases even if the 
resolved component is only taken into account on a LO basis, which in turn implies 
that our NLO corrections should already be sufficient to shed light on the question of 
general perturbative stability raised above. We also mention in this context
that our results for the NLO corrections to the direct hard subprocess cross sections
will help to obtain or check those for the resolved ones as the abelian ('QED-like') 
parts of the two are the same. 
The task amounts then to compute the matrix elements corresponding to the following $2\rightarrow 3$ processes
\begin{eqnarray}
\gamma q\rightarrow q+ X  \,\,\,\,\,\,\,\,
\gamma q\rightarrow g+ X \,\,\,\,\,\,\,\,
\gamma g\rightarrow q+ X  \\
\gamma g\rightarrow g+ X \,\,\,\,\,\,\,\,
\gamma q\rightarrow \bar{q}+ X \,\,\,\,\,\,\,\,
\gamma q\rightarrow q'+ X \nonumber 
 \end{eqnarray}
and the virtual corrections for the first three ones.

 They have been obtained from the original computation of the matrix elements contributing to polarized prompt photon production in \cite{wernerlionel} crossed to give the required ones. They are computed in dimensional regularization within the HVBM scheme \cite{hbvm} which provides a consistent implementation of $\gamma^5$. It should be noticed that the same matrix elements are the ones appearing at NLO in the case of photoproduction of either two jets or two charged hadrons, which has also been proposed to extract the polarized gluon distribution  from fixed target experiments \cite{bravar}. This computation is also a first step towards those interesting results. 

The phase space integration over the unobserved partons, the most complicated task, is done in the canonical way  and a precise description of the method can be found in the literature \cite{ellis}.
The infrared poles coming from the integrated $2\rightarrow 3$ processes are cancelled when the virtual corrections are added, and the remaining collinear poles are factorized in the corresponding distributions, including the photonic ones.
 \begin{figure}[htb]
\begin{center}
\mbox{\kern-1cm
\epsfig{file=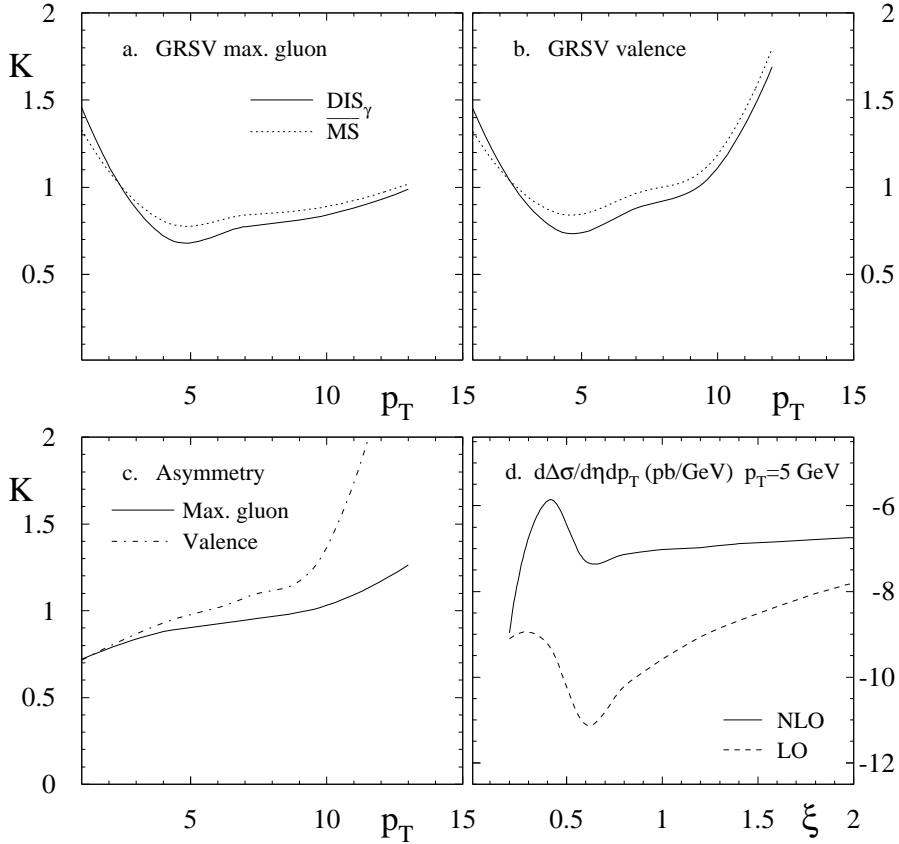,width=13.0truecm,angle=0}}
\caption{$K$-factors for polarized photoproduction of hadrons}
\end{center}
\end{figure}
In fig. 4 we show the results for the $K$-factor computed in the kinematical region of HERA more propitious to give information about $\Delta g$, i.e,   $\eta=-1$, as a function of the transverse momentum of the hadron. There are mainly two uncertanties which are shown in the plots: The one stated above about the fact that only the direct contribution has been computed at NLO, and then there is a dependence on the factorization scheme chosen for polarized parton distribution in the {\it photon}. In this case we show results in the most used schemes: $\overline{MS}$ and $DIS_\gamma$. As can be seen there is only a  mild dependence on them due to the fact that the direct contribution is much larger than the resolved one  in this kinematical region, and then  a large dependence of the {\it resolved} one on the scheme, to cancel a possible large  one from the direct part, cannot be expected.
The second uncertainty comes from the fact stated in the last section: The dominant polarized gluon distribution is not well constrained by the fits and the LO and NLO parametrizations can largely differ beyond the expectations from    perturbative stability. As a result, it turns out that the $K$-factor could be dominated by the ratio $\Delta G^{NLO}/\Delta G^{LO}$ instead of the hard cross section. In order to analyze it we compute the $K$-factor using two different polarized parametrizations \cite{grsv}: the usual GRSV Valence (fig. 4b) and a similar set where the polarized gluon distribution is assumed to be equal to the unpolarized one at a very small scale (${\cal{O}}(0.3\, $GeV$^2$)), and then, perturbatively stable by construction (fig. 2a). As can be observed the results are very similar in the region of $p_T<10$ GeV and differ only for larger values of $p_T$ which correspond to the kinematical region where the cross section becomes negligible (the end of the phase space for this process). The $K$-factor is found to be moderate for the polarized cross section in the region of $p_T>3$ GeV, and the situation is even better for the asymmetry where a cancellation between unpolarized and polarized $K$-factors occurs, as shown in fig. 4.c, where the unpolarized cross section at NLO has been computed  adding both direct and resolved contributions.

In fig. 4.d we show the dependence of the direct part of the polarized cross section on the arbitrary scales at a value of $p_T=5$ GeV and the same   rapidity as before, at both LO and NLO. In this case all the scales are assumed to be equal to $\xi p_T$, and the $\xi$ dependence is analyzed. As can be observed the NLO result is much more stable and its dependence on $\xi$ is smaller than 10\% for $\xi>0.5$, showing the relevance of the corrections.

\section{Conclusions}

 Performing a LO and NLO global analysis to both inclusive and semi-inclusive polarized deep inelastic data, we have found that the present semi-inclusive data can be consistently included in global analyses. These global fits show features similar to those coming from totally inclusive data, i.e., a poorly constrained gluon distribution and better determined valence densities, with the semi-inclusive data  introducing  very  small modifications in the  valence densities.
 Present semi-inclusive data alone fail to define a $\Delta d_V$ distribution consistent with those extracted from inclusive data; consequently,  the corresponding sets cannot reproduce the inclusive asymmetries for neutron targets.  However, ongoing semi-inclusive experiments using  $^3$He targets, or more accurate measurements on proton and deuteron targets, can reverse this situation and provide an enhanced perspective of the spin structure of the nucleon.

We have also presented for the first time the NLO corrections to the direct component  of the photoproduction of charged hadrons, which is a process specially suited to give information on th polarized gluon distributions. It was found that the observable is perturbatively stable in the region of $p_T>3$ GeV for the HERA kinematics, validating the existing LO analysis.
Both semi-inclusive observables, usually not taken into account in the analysis for the unpolarized case due to the large variety of available data, will play a fundamental role in order to increase our knowledge about the polarized structure of the hadrons.
We warmly acknowledge  R. Sassot and  W. Vogelsang for fruitful collaborations
in the first and second part of this work respectively.
  This work of  was partially supported by the World Laboratory.
 
 Este trabajo esta dedicado a la memoria de Ernesto "Che" Guevara.

\end{document}